\pdfoutput=1
\documentclass[aps,prl,twocolumn,superscriptaddress,amsfont,graphicx,nofootinbib,preprintnumbers]{revtex4-2}
\usepackage{graphicx}
\usepackage{amsmath}
\usepackage{slashed}
\usepackage{color,graphicx,epsfig}
\usepackage{ifpdf}
\usepackage{bm}
\usepackage{color}
\usepackage[english]{babel}
\usepackage{graphicx}
\usepackage{amsfonts}
\usepackage{amssymb}
\usepackage{braket}
\usepackage{hyperref}
\usepackage{xspace}
\usepackage{enumerate}

\newcommand{\gsim}{\;\rlap{\lower 3.5 pt \hbox{$\mathchar \sim$}} \raise 1pt \hbox {$>$}\;}
\newcommand{\lsim}{\;\rlap{\lower 3.5 pt \hbox{$\mathchar \sim$}} \raise 1pt \hbox {$<$}\;}

\def\LEe{L_{\mu e}}
\def\me{m_{e}}
\def\mE{m_{\mu}}
\def\mesq{m_{e}^2}
\def\mEsq{m_{\mu}^2}
\def\ne{n_{e}}
\def\nE{n_{\mu}}

\begin{document}

\preprint{ALBERTA-THY-3-25, CERN-TH-2025-190, TUM-HEP-1574/25}

\title{High-Energy Evolution of Power-Suppressed Amplitudes}

\author{Maximilian Delto}
\affiliation{Theoretical Physics Department, CERN,  CH-1211 Geneva 23, Switzerland}

\author{Alexander Penin}
\affiliation{Department of Physics, University of Alberta,
Edmonton AB T6G 2E1, Canada}

\author{Lorenzo Tancredi}
\affiliation{TUM School of Natural Sciences, Physik Department, James-Franck-Straße 1,
Technische Universität München, D--85748 Garching, Germany}

\preprint{}

\begin{abstract}
We present a new class of evolution equations which govern the
high-energy behavior of power-suppressed scattering amplitudes.
The equations can be viewed as a renormalization group flow
with respect to the relevant effective field theory cutoff. A
distinct feature of the method is in the use of a
multidimensional cutoff to separate the relevant scales in
problems characterized by a complex factorization structure. By
adjusting the renormalization group variables to the geometry
of the effective theory modes, our method naturally  extends to
a broad spectrum of physical problems including massive,
massless, small, and wide angle scattering. We present
applications to the benchmark processes of electron-positron
forward annihilation and light quark mediated Higgs boson
production/decays.
\end{abstract}

\maketitle

The high-energy  behavior of scattering amplitudes is
significantly altered by quantum effects. It is among  the
early applications of  quantum field theory
\cite{Sudakov:1954sw,Gorshkov:1966ht,Cheng:1970xm,
Frolov:1970ij,Frenkel:1976bj} and has  become a classical
problem  playing a fundamental role in particle phenomenology.
The asymptotic behavior of amplitudes which are not suppressed
by the ratio of a characteristic infrared scale to the process
energy is by now well understood
\cite{Libby:1978qf,Mueller:1979ih,Collins:1980ih,Sen:1981sd,
Collins:1985ue,Sterman:1986aj,Korchemsky:1988hd}. However, the
significant progress of experimental measurements and
perturbative calculations in a persistent search for physics
beyond the standard model, has gradually shifted the focus of
theoretical studies to  power-suppressed contributions. Besides
their phenomenological relevance, power corrections reveal many
intriguing properties and pose one of the main challenges to
modern effective field theory methods. The central problem of
the analysis is the resummation of radiative  corrections
enhanced by a power of the logarithm of the large scale ratio
which breaks down the finite-order perturbation theory and
determines the asymptotic behavior of the amplitudes. Over the
last decade, a wide spectrum of  power corrections  has been
extensively analyzed within different frameworks (see
\cite{Laenen:2010uz,Penin:2014msa,
Melnikov:2016emg,Bonocore:2016awd,Liu:2017vkm,Boughezal:2016zws,
Moult:2018jjd,Liu:2018czl,Beneke:2018gvs,Ebert:2018gsn,
Bahjat-Abbas:2019fqa,Beneke:2019mua,Moult:2019uhz,Liu:2019oav,
Anastasiou:2020vkr,Liu:2020wbn,Penin:2019xql,Beneke:2022obx,
Liu:2021chn,Liu:2022ajh,AH:2020iki,Bell:2022ott,Bell:2024bxg,
Liu:2024tkc} for the examples), but only a handful of complete
closed-form all-order results for physical observables have
been obtained. The general structure of the renormalization
group flow turns out to be far more complex and often even the
relevant effective-theory modes differ from the leading power.
Hence, while the general concept of scale separation is still
at the core of any analysis, the classical {\em hard-jet-soft}
factorization \cite{Libby:1978qf} has to be significantly
advanced. The common approach based on soft-collinear effective
theory has to deal with the problem of end-point singularities,
which require a nontrivial case-specific refactorization of the
effective-theory building blocks
\cite{Beneke:2022obx,Bell:2022ott,Liu:2022ajh}. At the same
time, the direct  resummation of the relevant  Feynman diagrams
becomes  quite nontrivial beyond the leading logarithms
\cite{Anastasiou:2020vkr}.

In this Letter we present a solution to this problem by
introducing a new class of evolution equations for power
suppressed amplitudes. These equations describe the change of
the amplitudes under the variation of the effective-theory
cutoff and, therefore, can be understood as a Wilsonian
renormalization group flow. The main difference with respect to
the existing analysis is the use of more than one cutoff
dictated by the factorization structure and the geometry of the
effective-theory modes, which makes the renormalization group
flow multidimensional. The building blocks that form the system
of  evolution equations in general differ from and have more
complex multiscale structure than the standard set of hard, jet
and soft functions. The method is universally applicable to a
wide range of physical processes including massive, massless,
small, and wide-angle scattering and to power corrections
in mass and transverse momentum. We introduce the method by
deriving the electron-to-muon forward annihilation (or backward
scattering) amplitude through the next-to-leading logarithmic
approximation. This is a classical problem
\cite{Gorshkov:1966ht} and its completely distinct spectrum of
modes and factorization structure poses a severe challenge to
the soft-collinear effective theory approach
\cite{Bell:2022ott}. Moreover, the available two-loop result
for the electron–muon scattering amplitudes
\cite{Bonciani:2021okt}, obtained using infrared matching
\cite{Penin:2005kf,Penin:2005eh,Bonciani:2007eh}, cannot be
extended to the forward region because of its peculiar infrared
structure. For this reason,  we complete the two-loop analysis
of the annihilation process by presenting the full result for
the forward amplitudes. We then extend the renormalization
group analysis to the light-quark-mediated Higgs boson
production and decays \cite{Liu:2017vkm}.

We start with an outline of the main idea. First, asymptotic
expansion of Feynman integrals
\cite{Beneke:1997zp,Smirnov:2002pj} and gauge invariance are
used to reveal the factorization structure and the spectrum of
dynamical modes in a given kinematical regime. This enables
their effective theory description, often avoiding the tedious
construction of a complete effective action
\cite{Kuhn:1999nn,Penin:2016wiw,Gardi:2022khw,
terHoeve:2023ehm,Ma:2023hrt,Gardi:2024axt}.\footnote{Although
there is no one-to-one correspondence between expansion by
regions and the physical logarithmic momentum regions, the
former becomes instrumental for the determination of the
relevant modes at subleading power, see {\em e.g.}
\cite{Penin:2014msa,Penin:2016wiw}.} A set of cutoffs are
introduced which separate the scales and shape the phase space
of the effective-theory modes. We refrain from using the
dimensional regularization parameter as a cutoff as it is blind
to the geometry of the modes. The variation of the effective
theory amplitudes under the variation of the cutoffs is then
computed yielding a system of coupled evolution equations.
Solving the evolution equations and removing the auxiliary
cutoffs sums up the large logarithms of scale ratios and
provides the asymptotic behavior of the amplitudes. The
specific form of the differential operators, which result in a
closed system of the evolution equations with the corresponding
boundary conditions, is determined by the geometry of the
effective-theory modes. The resulting  equations are, in
general, of the second order, as dictated by the
two-dimensional nature of the infrared singularities of the
on-shell amplitudes.

\begin{figure}[t]
\begin{center}
\begin{tabular}{ccc}
\includegraphics[width=2.2cm]{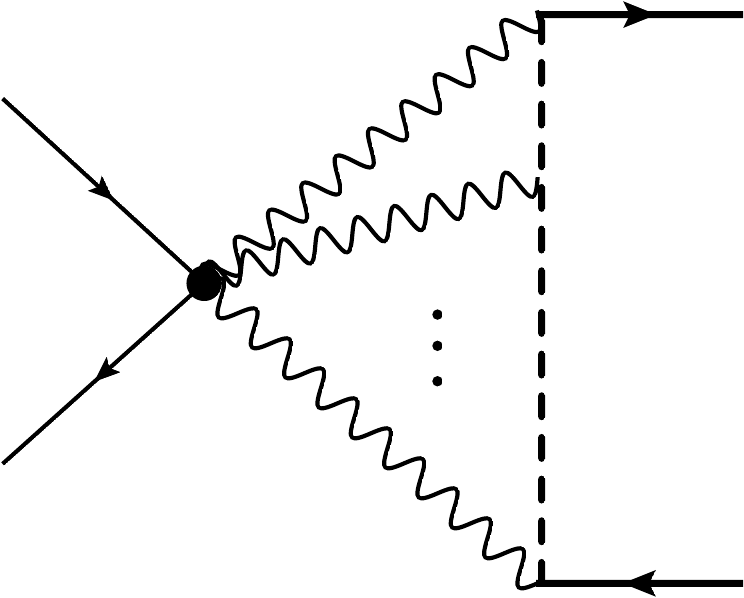}&
\hspace*{3mm}\includegraphics[width=2.2cm]{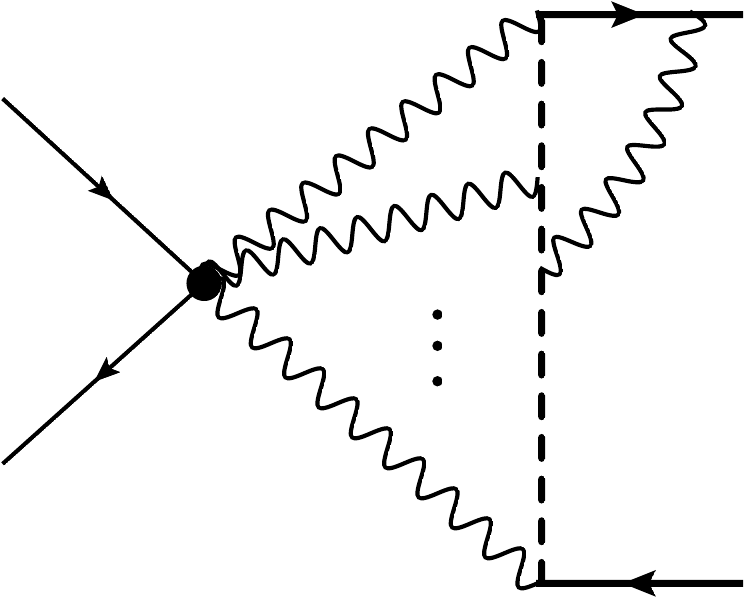}&
\hspace*{3mm}\includegraphics[width=2.2cm]{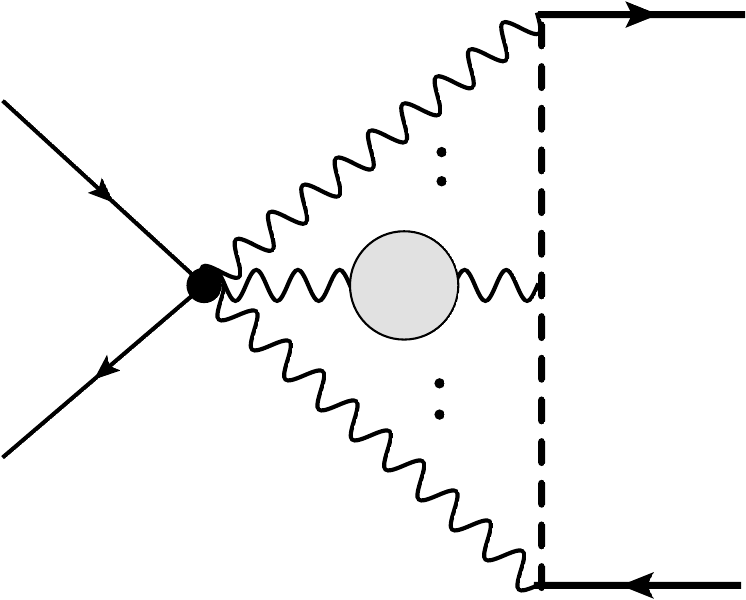}\\[2mm]
(a)&\hspace*{5mm}(b)&\hspace*{5mm}(c)\\
\end{tabular}
\end{center}
\caption{\label{fig::2} (a) The effective all-order  Feynman
diagrams representing  the leading logarithmic contribution.
(b) An example of topology not contributing in the
next-to-leading logarithmic approximation. (c) The effective
all-order  Feynman diagrams corresponding to the next-to-leading
logarithmic contribution.  The gray circle represents  the
one-loop single-logarithmic corrections to the annihilation
amplitude.}
\end{figure}

Let us now demonstrate how this program is realized for $e^+
e^- \to  \mu^+  \mu^-$ forward annihilation or, equivalently,
$e^- \mu^- \to  e^- \mu^-$ backward scattering. We choose the
reference frame in which, for vanishing electron mass, the electron
(positron) momentum has only one light-cone component $p^+$
($p^-$). In the high-energy limit the ratio of the fermion
masses $m_{e,\mu}$ to the total energy $\sqrt{s}$ is small. The
forward  regime is defined for small scattering angles
$\theta\lsim m_\mu^2/s$ when the total momentum transfer scales
as $(m_\mu^2-m_e^2)/\sqrt{s}$. Thus the contribution of the
forward region to the total annihilation cross section is
power-suppressed by the final state phase space. In the forward
regime the exponential suppression of the amplitude due to the
double-logarithmic Sudakov corrections associated with the soft
photon emission is absent but at the same time the amplitude
develops characteristic non-Sudakov radiative  corrections
enhanced by the second power of the large logarithm
$L=\ln(s/m_\mu^2)$ per each power of the fine structure
constant $\alpha$, associated with the soft fermion exchange.
To account for the factorization of mass singularities, we
write the amplitude of the process in the form ${\cal M}={\cal
Z}_{e\mu}A(z)\, {\cal M}_{\rm Born}+\ldots$, where $\alpha$ in
the Born amplitude is renormalized at the physical scale
$\sqrt{s}$ and  the factor ${\cal Z}_{e\mu}$ describes the
collinear renormalization of the external on-shell lines. In
the leading approximation it is given by  ${\cal
Z}_{e\mu}=\left({s/(m_e m_\mu)}\right)^{{\alpha\over
\pi}\gamma^{(1)}_f}$, where $\gamma^{(1)}_f=3/2$ is the
one-loop collinear anomalous dimension. The form factor $A(z)$
is a function  of  the variable $z={\alpha L^2/2\pi}$ and
accommodates the double logarithmic terms to all orders in
$\alpha$. These non-Sudakov double logarithms originate from
the planar multiloop annihilation diagrams
\cite{Gorshkov:1966ht}. Due to the absence of momentum
transfer, the electron propagator of each loop gets canceled,
while the muon propagators effectively become scalar and
sufficiently singular to produce the double logarithmic
corrections, see Fig.~\ref{fig::2}(a). The relevant effective
theory with hard scale $\sqrt{s}$ and soft scale $m_\mu$
involves soft on-shell muons with loop momenta $l_i$ and
propagators $i\pi \delta(l_i^2-m_\mu^2)$, and eikonal off-shell
transverse photons with propagators $g^{\perp}_{kl}/(2l_il_j)$.
It is sufficient for the diagrammatic calculation and
resummation of the double-logarithmic terms. We, however,
reformulate the problem in the spirit of renormalization-group
flow, and introduce an auxiliary ultraviolet cutoff
$\nu^+<\sqrt{s}$ on the plus-components of the light-cone
momenta and an infrared cutoff $m_\mu^2/\sqrt{s}<\nu^-<\nu^+$
on the minus-components. The form factor now becomes a function
of the cutoffs $A(\xi,\eta)$, where we introduce the normalized
logarithmic variables $\xi=\ln(\nu^+\sqrt{s}/m^2_\mu)/L$,
$\eta=-\ln(\nu^-/\sqrt{s})/L$.  The  full-theory amplitude is
then given by $A(z)=A(1,1)$ corresponding to the physical
cutoff $\nu^+=\sqrt{s}$, $\nu^-=m_\mu^2/\sqrt{s}$. The
double-logarithmic scaling of the $\ell$-loop diagrams requires
the (anti-)collinear $l_i^\perp\sim m_\mu$, $l_i^\pm\sim
m_\mu^2/\sqrt{s}$, $l_i^\mp\sim \sqrt{s}$ or the soft
$l_i^\perp \sim \,l_i^\pm\sim m_\mu$ scaling of the loop momenta
components and the hierarchy $l^+_\ell<\ldots l^+_1<\nu^+$,
$\nu^-<l^-_1<\ldots< l^-_\ell$. Hence, the second-order
derivative with respect to $\xi$ and $\eta$ captures the
ultraviolet and infrared singularities of the $l_1$ integral,
and  maps the $\ell$-loop into the $(\ell-1)$-loop effective
theory diagram. Thus, the change of the all-order leading
logarithmic form factor with the variation of the cutoffs is
proportional to the form factor itself. This yields the
following evolution equation
\begin{equation}
{\partial^2 A(\xi,\eta)\over \partial\xi\partial\eta}
=zA(\xi,\eta)\,,
\label{eq::evoleqLL}
\end{equation}
with  boundary conditions
\begin{equation}
A(\xi,0)=1\,,
\qquad \left.{\partial
A(\xi,\eta)\over \partial \eta}\right|_{\eta=\xi}=0\,.
\label{eq::bcll}
\end{equation}

\begin{figure}[t]
\begin{center}
\begin{tabular}{c}
\includegraphics[width=3.0cm]{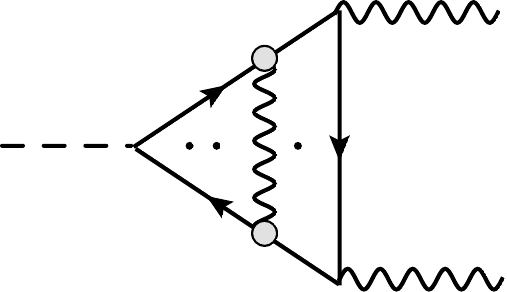}\\
\end{tabular}
\end{center}
\caption{\label{fig::3} The   Feynman diagram with an
arbitrary number of Sudakov gluon exchanges representing  the
leading logarithmic corrections to the light quark mediated
$H\to gg$ amplitude. The gray circles represent the
color-adjusted effective vertices. }
\end{figure}

\noindent
The first condition indicates that when the infrared cutoff
approaches the physical hard scale $\nu^-=\sqrt{s}$, the
effective-theory phase space shrinks to nothing and the
amplitude reduces to the Born result.  On the other hand, the
pinch condition on the pole of the soft muon propagator
$l^-_i>m_\mu^2/l^+_i$ sets a natural infrared cutoff
$m_\mu^2/\nu^+$  on the $l^-_1$ momentum component, which
means that  for $\nu^-\le m_\mu^2/\nu^+$ corresponding to
$\eta\ge \xi$ the amplitude does not depend on the infrared
cutoff $\nu^-$, which gives the second boundary condition. The
solution of Eqs.~(\ref{eq::evoleqLL},\,\ref{eq::bcll}) reads
\begin{equation}
A(\xi,\eta)=I_0\left(2(z\eta\xi)^{1/2}\right)
-{\eta\over \xi}I_2\left(2(z\eta\xi)^{1/2}\right)\,,
\label{eq::Ares}
\end{equation}
where $I_n(x)$ is the $n$th modified Bessel function.
This recovers the leading logarithmic result
\cite{Gorshkov:1966ht}
\begin{equation}
A(z)={I_1(2z^{1/2})\over z^{1/2}}=1+{z\over 2}+{z^2\over 12}
+\ldots\,,
\label{eq::A11}
\end{equation}
with $A(z)\sim {e^{2\sqrt{z}}\over 2\sqrt{\pi} z^{3/4}}$
at $z\to\infty$.

The analysis readily generalizes beyond the leading-logarithmic
order. First we note that in Fig.~\ref{fig::2}(a) every loop
can admit the forward kinematics mandatory to produce the
double logarithmic scaling. At the same time, radiative
corrections of the type exemplified in Fig.~\ref{fig::2}(b)
destroy this kinematics when the extra loop momentum exceeds
$m_\mu$ and, in this way, they reduce the power of the large
logarithm by two. Hence, the diagram Fig.~\ref{fig::2}(b) does
not contribute to the next-to-leading running, which is instead
completely determined by the diagrams of topology as in
Fig.~\ref{fig::2}(c), with the gray circle representing the
soft single-logarithmic terms in the one-loop
forward annihilation amplitude  with external momenta $l_i$ and
$l_{i+1}$. This simple factorization structure makes the
calculation of the amplitude dependence on the cutoff at this
order rather straightforward
\begin{equation}
\begin{split}
&\hspace*{-3mm}{1\over z}
{\partial^2 A(\xi,\eta)\over \partial\xi\partial\eta}
=\left(1-{\alpha L\over 2\pi}(\xi-\eta)
\right) A(\xi,\eta)\\
&\hspace*{-3mm} +z
\int_{0}^\eta{\rm d}\eta'\int_{\eta'}^\xi{\rm d}\xi'\,
\delta(\xi-\eta')A(\xi',\eta') +\delta(\xi)A(0,0)\,.
\label{eq::evoleqNLL} 
\end{split}
\end{equation}
A finite electron mass results in the reduction of the
double-logarithmic integration region, which  can be accounted
for in the next-to-leading logarithmic approximation by  a
redefinition of the double logarithmic variable to
$z={\alpha\over 2 \pi}{L'}^2$ with
\begin{equation}
L'= \ln\left(e^{\gamma^{(1)}_e}{s\over
m_\mu^2}\right)=L+\gamma^{(1)}_e\,,
\label{eq::Lprime}
\end{equation}
where
\begin{equation}
\gamma^{(1)}_e=-{m_e^2\over m_\mu^2-m_e^2}
\ln\left({m_\mu^2\over m_e^2}\right)\,. 
\label{eq::gammam}
\end{equation}
In Eq.~(\ref{eq::evoleqNLL}) the function
\begin{equation}
\delta(\xi)={\alpha(e^{{L\over 2}\xi}m_\mu)
\over \alpha(m_\mu)}-1\,,
\label{eq::del}
\end{equation}
accommodates the leading effect of the fine structure constant
running, with renormalization scale of $\alpha$ in the double
logarithmic variable $z$  set to $m_\mu$. A rather complex
integro-differential Eq.~(\ref{eq::evoleqNLL}) can in principle
be solved numerically retaining the all-order logarithmic
result for the function $\delta(\xi,\eta)$.
However, if we work in the strict next-to-leading logarithmic
approximation {\it i.e.} only keep the one-loop term, Eq.~(\ref{eq::evoleqNLL})
reduces to a second-order linear partial differential equation
\begin{equation}
\begin{split}
&{\partial^2 A(\xi,\eta)\over \partial\xi\partial\eta}
=z\left[1-{\alpha L\over 2\pi}(\xi-\eta)\right.
\left.\left({\beta_0\over 2}+1\right)\right]A(\xi,\eta)\,,
\label{eq::evoleqNLLexp}
\end{split}
\end{equation}
where $\beta_0=-{4\over 3}n_l$ is the QED beta function with
$n_l=2$ active flavors at the scale $\sqrt{s}$, which can be
solved perturbatively about the leading-logarithmic result
Eq.~(\ref{eq::Ares}). The first boundary condition is now
modified by matching to the one-loop amplitude
\begin{equation}
A(\xi,0)=1-{\alpha L\over 2\pi}\,,
\label{eq::bcnll}
\end{equation}
while the second kinematical boundary condition does not
change. The final next-to-leading logarithmic  result for the
form factor reads
\begin{eqnarray}
\hspace*{-3mm}A(z)\!\!&=&\!\!
\bigg(1-{\alpha L\over 2\pi}\bigg){I_1(2z^{1/2})\over z^{1/2}}
-\!{\alpha L\over 2\pi}\left({\beta_0\over 2 }+1\right)\!k(z)\,,
\label{eq::MNLL}
\end{eqnarray}
where 
\begin{eqnarray}
k(z)&=&z\int_{0}^1{\rm d}\eta\int_{\eta}^1{\rm d}\xi
(\xi-\eta)A(\xi,\eta)A(1-\eta,1-\xi)
\nonumber \\
&=&{z\over 6}+{z^2\over 20}+{29\over 5040}z^3+\ldots\,,
\label{eq::kz}
\end{eqnarray}
where $A(\xi,\eta)$ is given by Eq.~(\ref{eq::Ares}), with the
asymptotic behavior $k(z)\sim {e^{2\sqrt{z}}\over 4\sqrt{z}}$
at $z\to\infty$.  This calculation  provides  the ${\cal
O}(\alpha^2)$  virtual corrections to the cross  section of
$e^+ e^- \to \mu^+  \mu^-$ annihilation in forward kinematics
through the next-to-leading logarithms. Writing
$d\sigma=\sum_{n=0}^\infty({\alpha\over
2\pi})^{n}d\sigma^{(n)}$ for the one and two-loop terms we get
\begin{eqnarray}
{d\sigma^{(1)}\over d\sigma_{\rm Born}}&=&
L^2+\left(2\gamma^{(1)}_e+4+{4\over 3}n_l\right) L\,,
\label{eq::sigmaNLO}\\
{d\sigma^{(2)}\over d\sigma_{\rm Born}}&=&
{5\over 12}L^4
+\left(\frac{5}{3}\gamma^{(1)}_e +{11\over 3}  +{14\over 9}n_l\right)L^3\,,
\label{eq::sigmaNNLO}
\end{eqnarray}
where we retain the full dependence on the mass ratio in the 
coefficients and the coupling constant in the Born cross section
is renormalized at $m_\mu$. While
Eqs.~(\ref{eq::sigmaNLO},\ref{eq::sigmaNNLO}) are finite, the
virtual corrections to the cross section become infrared
divergent starting from the next-to-next-to-leading logarithmic
approximation. To get the  physical finite contribution to the
cross section from the forward region, the real emission of
soft photons with characteristic energy $\varepsilon\lsim
m_\mu^2/\sqrt{s}$ should be added, which factors out with
respect to the virtual corrections discussed above. We note
that for $m_e = m_\mu$  the differential cross section for
strictly forward annihilation is infrared finite beyond the
logarithmic approximation since the soft emission vanishes
identically along with the momentum transfer. To verify this
analysis, we have analytically evaluated all two-loop terms
surviving the high-energy limit for $m_e\ll m_\mu$ and $m_e =
m_\mu$, building upon earlier work by some of
us~\cite{Delto:2023kqv}. This allowed us to confirm the leading
and next-to-leading logarithmic contribution given above.  The
full result for the forward amplitudes in the limit of the
small electron mass is given in \cite{Supp}.  We should note
that with the photon mass used as an infrared regulator in the
derivation of Eq.~(\ref{eq::evoleqNLL})  the  soft photon
emission can be unambiguously separated and the coefficients of
Eqs.~(\ref{eq::sigmaNLO},\ref{eq::sigmaNNLO}) do not depend on
the infrared subtraction scheme. At the same time if instead
the dimensional regularization is used, the (finite)
coefficients of the next-to-leading logarithms do depend on the
subtraction scheme and Eq.~(\ref{eq::sigmaNNLO}) corresponds to
the one of \cite{Becher:2009kw,Ferroglia:2009ep}, 
see \cite{Supp}.

Let us show how the method  extends to problems with quite a
different dynamics. We start with the light-quark mediated
Higgs boson production and decays. A detailed analysis of the
logarithmic corrections to the corresponding $ggH$ amplitude
through the next-to-leading logarithmic approximation can be
found in \cite{Liu:2017vkm,Anastasiou:2020vkr}. The origin of
the  double logarithms in this case is quite different and
involves a single chirality flipping soft fermion exchange
accompanied by all-order  {\it Sudakov} soft on-shell gluons,
see Fig.~\ref{fig::3}.   Nevertheless, we can apply the above
formalism  to derive the equations governing the
renormalization group evolution  of the amplitude. The hard and
soft scales in the problem are given by the Higgs boson mass
$m_H$ and a light-quark mass $m_q\ll m_H$, respectively. We
introduce independent infrared  cutoffs $m_q^2/m_H<
\nu^\pm<m_H$  on the light-cone momentum components with the
corresponding  normalized logarithmic variables
$\xi=-\ln(\nu^+/m_H)/L$, $\eta=-\ln(\nu^-/m_H)/L$, where
$L=\ln(m_H^2/m_q^2)$. It is convenient to define the form
factor $A(z)$  normalized to $1$ for the leading order one-loop
mass-suppressed amplitude, ${\cal M}={\cal Z}_{g}A(z){\cal
M}_{\rm LO}+\ldots$, where $z={\alpha_s L^2/2\pi}$, $\alpha_s$
is the strong coupling constant $\alpha_s(\mu)$ renormalized at
the scale $\mu=m_q$, and we have separated the standard Sudakov
factor ${\cal Z}_{g}$ for the external on-shell gluon lines,
which incorporates all the infrared divergences of the
amplitude. In the effective theory, the form factor becomes a
function $A(\xi,\eta)$ of  the cutoff  with $A(z)=A(1,1)$. By
using the results of \cite{Liu:2017vkm,Anastasiou:2020vkr} on
the factorization of the logarithmic regions,  we  compute its
variation, which in the leading-logarithmic approximation
yields the  evolution equation
\begin{equation}
{\partial^2 A(\xi,\eta)\over \partial\xi\partial\eta}
=2F(\xi,\eta,0)\,,
\label{eq::evoleqHiggs}
\end{equation}
where $F(\xi,\eta,\zeta)$  is the  off-shell  Sudakov form
factor normalized to $F(\xi,\eta,\zeta)|_{z=0}=1$. It is
defined for the (anti)quark square momenta  equal to  $\nu^\pm
m_H$,  and is a function of $\zeta=-\ln(\nu^2/m_H^2)/L$ with
$\nu^\pm< \nu< m_H$ being an ordinary hard momentum cutoff. The
dependence of the form factor on the hard cutoff is well
understood  \cite{Korchemsky:1988hd} and is determined by the
equation
\begin{equation}
{\partial \ln\!F(\xi,\eta,\zeta)\over \partial\zeta}
=4\tilde \gamma_{\rm cusp}^{(1)}z(\xi+\eta-2\zeta)\,.
\label{eq::evoleqSud}
\end{equation}
Here $\tilde\gamma_{\rm cusp}^{(1)}=-C_F+C_A$, with $C_A=N_c$,
$C_F={N_c^2-1\over 2N_c}$, is the difference of the one-loop
cusp anomalous dimension of the light-like Wilson lines in the
fundamental and adjoint representation of the $SU(N_c)$ color
group.  This subtraction accounts for the color charge
variation along the light-cone Wilson lines in the process of
soft-quark emission, which is the physical origin of the
double-logarithmic corrections to the amplitude. The evolution
of the form factor is determined by the collinear, soft and
ultrasoft modes with the infrared scales set by the effective
theory cutoffs $\nu^\pm$. Thus, the form factor has a
multiscale structure and its renormalization group analysis
requires further scale separation  and introduction of
additional cutoffs \cite{Korchemsky:1988hd}, which disappear
from the final evolution equation. Due to the scale hierarchy
these auxiliary  cutoffs do not interfere with the
renormalization group parameters $\xi$, $\eta$. The solution of
Eq.~(\ref{eq::evoleqSud}) for $F(\zeta,\zeta,\zeta)=1$ is
$F(\xi,\eta,\zeta)=e^{4\tilde\gamma_{\rm
cusp}^{(1)}z(\xi-\zeta)(\eta-\zeta)}$. Then
Eq.~(\ref{eq::evoleqHiggs})  can be integrated with the
boundary conditions
\begin{equation}
A(\xi,0)=0\,, \qquad
\left.{\partial A(\xi,\eta)\over \partial \eta}\right|_{\eta=1-\xi}=0
\label{eq::bcHiggs}
\end{equation}
which can be derived in the same way as Eq.~(\ref{eq::bcll}).
The solution reads
\begin{equation}
A(\xi,\eta)=2\int_0^\eta {\rm d}\eta'
\int_{1-\xi}^{1-\eta'}{\rm d}\xi'
e^{4\tilde\gamma_{\rm cusp}^{(1)}z\eta'\xi'}\,,
\label{eq::solHiggs}
\end{equation}
and reproduces the well known leading logarithmic result
\cite{Liu:2017vkm,Liu:2018czl}
\begin{equation}
A(z)={}_2F_2\left(1,1;{3/2},2;{\tilde\gamma_{\rm cusp}^{(1)}z}\right)\,,
\label{eq::resHiggs}
\end{equation}
where  ${}_2F_2\left(1,1;{3/2},2;{x}\right)=1+{x\over 3}
+\ldots$ is the hypergeometric function with the asymptotic
behavior ${}_2F_2\left(1,1;{3/2},2;{x}\right)\sim
\sqrt{\pi\over x} {e^{x}\over 2}$ at $x\to\infty$. As for
Eq.~(\ref{eq::evoleqNLL}), in the next-to-leading logarithmic
approximation one includes the collinear renormalization of the
quark fields and the renormalization group running of the
strong coupling constant into the leading order equations. This
results in the additional factor
$Z_q(\xi)Z_q(\eta)=e^{{\alpha_s L\over
4\pi}\gamma_q^{(1)}(3\xi+3\eta-2)}$, $\gamma_q^{(1)}=3C_F/2$,
on the right hand side of Eq.~(\ref{eq::evoleqHiggs}) while the
equation for the form factor becomes
\begin{eqnarray}
{\partial \ln\!F(\xi,\eta,\zeta)\over \partial\zeta}
&=&4\tilde \gamma_{\rm cusp}^{(1)}z
\int_\zeta^{\xi+\eta-\zeta}{\rm d}\zeta'
\alpha_s(e^{-L\zeta'/2}m_H)
\nonumber\\
&+&{L\alpha_s\over 2\pi}\gamma_q^{(1)}\,.
\label{eq::evoleqSudnll}
\end{eqnarray}
The above system of coupled equations can be solved numerically
or perturbatively about the leading logarithmic solution
Eq.~(\ref{eq::solHiggs}), after expanding the collinear factor
and the running coupling to the first order in $\alpha_s$. The
expansion reproduces the analytic all-order result obtained
within the diagrammatic approach \cite{Anastasiou:2020vkr} and
the numerical three-loop result \cite{Czakon:2020vql}.

Unlike the existing approaches based on dimensional
regularization the method does not require additional rapidity
regulators or the refactorization, and the above analysis
naturally extends to the massive form factors
\cite{Penin:2014msa,Liu:2017vkm}, where the renormalization
group description is yet to be formulated at subleading power
in the small mass expansion. Finally we note that the method
can be generalized to the subleading power dynamics involving
Glauber modes (see {\it e.g. } \cite{Penin:2019xql}). In this
case the renormalization group variables are to be associated
with the infrared   cutoff on the light-cone components  and an
ultraviolet  cutoff  on the transverse components in momentum
space.

In conclusion, we have developed a new approach to the analysis
of the high-energy asymptotic behavior of scattering amplitudes
at subleading power. It is based on multidimensional
renormalization-group flow described by (a system of) partial
differential equations. It provides a powerful and adaptive
tool applicable to  a  wide range of  processes and kinematical
regimes with a diverse spectrum of physical modes, from soft
fermions to Glauber gauge bosons, which  pose a serious
challenge to existing techniques. We also present the ${\cal
O}(\alpha^2)$ radiative corrections to the  amplitude of $e^+
e^- \to  \mu^+  \mu^-$ annihilation at high energy in the
forward limit, which completes the QED analysis of the process
at this order.

\begin{acknowledgments}
\emph{Acknowledgments:}
This research was supported in part by the Excellence Cluster
ORIGINS funded by the Deutsche Forschungsgemeinschaft (DFG,
German Research Foundation) under Germany’s Excellence Strategy
– EXC-2094-390783311 (M.D. and L.T.), in part by the European
Research Council (ERC) under the European Union’s research and
innovation programme grant agreements 949279 (ERC Starting
Grant HighPHun) (M.D. and L.T.) and 101044599 (ERC Consolidator
Grant JANUS) (M.D.) and in part by NSERC and by the Perimeter
Institute for Theoretical Physics (A.P.). Views and opinions
expressed are however those of the authors only and do not
necessarily reflect those of the European Union or the European
Research Council Executive Agency. Neither the European Union
nor the granting authority can be held responsible for them. We
are especially thankful to the Munich Institute for Astro-,
Particle and BioPhysics (MIAPbP), funded by the DFG under
Germany’s Excellence Strategy – EXC-2094-390783311, where part
of this work was completed.

\end{acknowledgments}

\onecolumngrid
\allowdisplaybreaks
\setcounter{secnumdepth}{2}

\newpage

\section*{Supplemental Material}
We decompose the $e^+ e^- \to \mu^+ \mu^-$ forward-annihilation amplitude in terms of two form factors
\begin{align}
    \label{eqn:amp-FF-decomposition}
    \mathcal{A}\ = -\frac{i}{s} \left[ \left( \overline{V}_e \gamma^{\rho} U_e \right) \left( \overline{U}_\mu \gamma_{\rho}   \, V_\mu \right) \mathcal{F} + \frac{m_e}{m_\mu} \left( \overline{V}_e U_e \right) \left( \overline{U}_\mu  \, V_\mu \right) \widetilde{\mathcal{F}} \right] \,,
\end{align}
which are normalized to scale as $\mathcal{O}(1)$ in $\mesq \ll \mEsq \ll s$.
From the decomposition in Eq.~(\ref{eqn:amp-FF-decomposition}) we can see
that the contribution of $\widetilde{\mathcal{F}}$ is suppressed by
one power of $m_e$ and can be neglected in our approximation.
In what follows we will focus on form factor $\mathcal{F}$,
which is, in fact, the projection onto the tree-level amplitude.
Adapting the projection method of \cite{Peraro:2020sfm},
we work in the so-called ’t~Hooft-Veltman scheme \cite{tHooft:1972tcz}.
We carry out multiplicative UV renormalization in the on-shell scheme,
\begin{align}
    \label{eqn:FF-ren-def}
    \mathcal{F}_{\mathrm{ren}} = Z_{2,e}^{\text{OS}} \, Z_{2,\mu}^{\text{OS}} \, \mathcal{F}_{\mathrm{b}}\!\left[m_{\mathrm{b},e}(m_{e}),m_{\mathrm{b},\mu}(m_{\mu}),\alpha_{\mathrm{b}}(\alpha)\right]\,,
\end{align}
at a scale $\mu^2=\me \mE$. After UV renormalization we are
left with one-loop exact infrared (IR) poles~\cite{Yennie:1961ad}
\begin{align}
    \label{eqn:definition-IR-subtraction}
    \mathcal{F}_{\mathrm{fin}} = \exp \left[ - \frac{\alpha}{2\pi} \frac{z_{\text{IR}}}{\epsilon}  \right] \mathcal{F}_{\mathrm{ren}}\,,
\end{align}
where $\mathcal{F}_{i,\mathrm{fin}}$ are finite in four dimensions
and $z_{\textrm{IR}}$ denotes the  anomalous dimension that governs
the  soft singularities of the amplitude to all-orders
through exponentiation~\cite{Becher:2009kw,Ferroglia:2009ep}.
For $\mesq \ll \mEsq \ll s$, we find
\begin{align}
\label{eqn:ZIR-EM}
z_{\text{IR}} = {} & -2\pi i -2 + \LEe \,.
\end{align}
We note that checking the cancellation of UV and IR $1/\epsilon$ poles constitutes a non-trivial check of our results. We expand
\begin{align}
    \label{eqn:FF-fin-res-exp}
    \mathcal{F}_{\text{fin}}= 1 + \sum_{l=1}^{2}\left({\alpha\over 2\pi}\right)^{l} \sum_{\substack{a,b=0 \\ a+b \le l}}^{l} (\ne)^a (\nE)^b \left[  \mathcal{F}_{\text{fin}}^{(l,a,b,\text{re})} + i \pi \mathcal{F}_{\text{fin}}^{(l,a,b,\text{im})} \right] \,,
\end{align}
where $n_e$($n_\mu$) denotes the number massive electrons (muons). The coefficients in Eq.~(\ref{eqn:FF-ren-res-exp}) constitute one of the main results of our work; we will discuss technical details of their computation in a forthcoming publication. Using abbreviations $L=\ln(s/m_{\mu}^2)$ and $\LEe=\ln(m_\mu^2/m_e^2)$ of the main text, the coefficients read
\begingroup
\allowdisplaybreaks
\begin{alignat}{4}
    \label{eqn:FF-coeff-res-1L}
    & \mathcal{F}_{\text{fin}}^{(1,0,0,\text{re})} ={} && \frac{1}{2}L^2 + 2 L + \bigg[ \frac{3}{2} \LEe -5  + \frac{7\pi^2}{6}    \bigg] \,, \qquad
    && \mathcal{F}_{\text{fin}}^{(1,0,0,\text{im})} ={} && L + \bigg[ \LEe -2 \bigg] \,, \nonumber \\
    & \mathcal{F}_{\text{fin}}^{(1,1,0,\text{re})} ={}&& \frac{2}{3} L + \bigg[ \frac{2}{3} \LEe - \frac{10}{9} \bigg] \,, \qquad
    && \mathcal{F}_{\text{fin}}^{(1,1,0,\text{im})} ={} && -\frac{2}{3} \,, \\
    & \mathcal{F}_{\text{fin}}^{(1,0,1,\text{re})} ={}&& \frac{2}{3} L  - \frac{10}{9}  \,, \qquad
    && \mathcal{F}_{\text{fin}}^{(1,0,1,\text{im})} ={} && -\frac{2}{3} \,, \nonumber
\end{alignat}
\endgroup
at one loop, and
\begingroup
\allowdisplaybreaks
\begin{align}
    \label{eqn:FF-coeff-res-2L-A}
    \mathcal{F}_{\text{fin}}^{(2,0,0,\text{re})} ={}& \frac1{12}L^4 + \frac{5}{6} L^3
    + L^2 \bigg[ \frac{3\LEe}{4} - \frac1{2} - \frac{2\pi^2}{3}  \bigg]
    - L \bigg[ \LEe (\pi^2-3) + \frac{11}{4} -\frac{8\pi^2}{3} - 10 \zeta_3 \bigg]   \nonumber \\
    {}& - \bigg[ \LEe^2 \left( \frac{\pi^2}{2} -\frac{9}{8} \right) + \LEe \left(\frac{57}{8}-\frac{13\pi^2}{4} -6 \zeta_3  \right)   \nonumber \\
    {}& -29 + \frac{23\pi^2}{4} + \frac{16\pi^4}{45} - 6 \pi^2 \ln(2) + \frac{4}{3} \pi^2 \ln^2(2)
    - \frac{4}{3}\ln^4(2) - 32\, \text{Li}_4(1/2) + 30 \zeta_3 \bigg] \,,\nonumber \\
    \mathcal{F}_{\text{fin}}^{(2,0,0,\text{im})} ={}& \frac{2}{3} L^3 + L^2 \bigg[ \frac{\LEe}{2} + \frac{1}{2} \bigg]
    + L \bigg[ \frac{7\LEe}{2}  + \pi^2 - 3 \bigg] + \bigg[  \frac{3\LEe^2}{2}
    + \LEe \left( \frac{7\pi^2}{6} - 8 \right)  \nonumber \\
    {}& -\frac{13}{4} - \frac{11\pi^2}{3} +2 \pi^2 \ln(2) - 11 \zeta_3 \bigg] \,,  \nonumber \\
    \mathcal{F}_{\text{fin}}^{(2,1,0,\text{re})} ={}& \frac{4}{9} L^3 +L^2 \bigg[ \frac{2\LEe}{3} + \frac{5}{9} \bigg]
    + L \bigg[ \frac{11\LEe}{3} - \frac{64}{9} + \frac{4\pi^2}{3} \bigg] - \bigg[ \frac{7}{36} \LEe^3  \\
    {}&  - \frac{23}{9}\LEe^2  + \LEe \bigg( \frac{649}{54}  - \frac{35\pi^2}{18} \bigg)
    -\frac{1493}{54} + \frac{122\pi^2}{27} - \frac{4}{3} \zeta_3 \bigg] \,, \nonumber \\
    \mathcal{F}_{\text{fin}}^{(2,1,0,\text{im})} ={}& \frac{2}{3} L^2 + L \bigg[ 2 \LEe - \frac{50}{9} \bigg]
    + \bigg[ \frac{3}{2} \LEe^2 - 7 \LEe + \frac{304}{27} -\frac{\pi^2}{3} \bigg] \,, \nonumber \\
    \mathcal{F}_{\text{fin}}^{(2,0,1,\text{re})} ={}& \frac{4}{9} L^3 + \frac{5}{9} L^2 + L \bigg[ \LEe - \frac{64}{9} + \frac{4\pi^2}{3}  \bigg] \nonumber \\
    {}& - \bigg[ \frac{\LEe^3}{12}  - \frac{\LEe^2}{6} + \LEe \bigg( \frac{5}{3} - \frac{11\pi^2}{18} \bigg)
    -\frac{13297}{324} + \frac{145\pi^2}{27} -\frac{4}{3} \zeta_3 \bigg] \,, \nonumber \\
    \mathcal{F}_{\text{fin}}^{(2,0,1,\text{im})} ={}& \frac{2}{3} L^2 + L \bigg[ \frac{2\LEe}{3} - \frac{50}{9} \bigg]
    + \bigg[ \frac{\LEe^2}{6} - \frac{19}{9} \LEe + \frac{304}{27} -\frac{\pi^2}{3} \bigg] \,, \nonumber
\end{align}
\endgroup
as well as
\begin{alignat}{4}
\label{eqn:FF-coeff-res-2L-B}
& \mathcal{F}_{\text{fin}}^{(2,1,1,\text{re})} ={}&& \frac{8}{9} L^2 + L \bigg[ \frac{8}{9} \LEe - \frac{80}{27} \bigg] - \bigg[ \frac{40}{27} \LEe - \frac{200}{81} + \frac{8\pi^2}{9} \bigg] \,, \quad
&& \mathcal{F}_{\text{fin}}^{(2,1,1,\text{im})} ={}&&  - \frac{16}{9} L - \bigg[ \frac{8}{9} \LEe - \frac{80}{27}  \bigg] \,, \nonumber \\
& \mathcal{F}_{\text{fin}}^{(2,2,0,\text{re})} ={}&& \frac{4}{9} L^2 + L \bigg[ \frac{8}{9} \LEe - \frac{40}{27} \bigg] + \bigg[ \frac{4}{9} \LEe^2 -\frac{40}{27} \LEe + \frac{100}{81} - \frac{4\pi^2}{9} \bigg] \,, \quad
&& \mathcal{F}_{\text{fin}}^{(2,2,0,\text{im})} ={}&& - \frac{8}{9} L - \bigg[ \frac{8}{9}  \LEe -\frac{40}{27}\bigg] \,, \\
& \mathcal{F}_{\text{fin}}^{(2,0,2,\text{re})} ={}&& \frac{4}{9} L^2 - \frac{40}{27} L  + \bigg[  \frac{100}{81} - \frac{4\pi^2}{9} \bigg] \,, \quad
&& \mathcal{F}_{\text{fin}}^{(2,0,2,\text{im})} ={}&& - \frac{8}{9} L +  \frac{40}{27} \,, \nonumber
\end{alignat}
at two loop. We note that our one-loop result agrees with the one in \cite{Bell:2022ott} after crossing from forward annihilation to backward scattering and adapting to the kinematical region $\mesq \sim \mEsq \ll s$.

In the main text, we have presented virtual corrections to the cross section for $e^+e^- \to \mu^+\mu^-$ through NLL in Eqs.~(\ref{eq::sigmaNLO},\ref{eq::sigmaNNLO}). In terms of the form factor $\mathcal{F}$, they read
\begin{align}
    \frac{\mathrm{d}\sigma^{(1)}}{\mathrm{d}\sigma_{\mathrm{Born}}} ={} \frac{2 \mathcal{M}_{0,1}}{\mathcal{M}_{0,0}} \,, ~~~~ \frac{\mathrm{d}\sigma^{(2)}}{\mathrm{d}\sigma_{\mathrm{Born}}} = \frac{2 \mathcal{M}_{0,2} + \mathcal{M}_{1,1}}{\mathcal{M}_{0,0}}  \,,
\end{align}
where
\begin{align}
 \mathcal{M}_{\left(l,r\right)} =  8\, \Re \left\{ \left[ \mathcal{F}_{\text{fin}}^{(l)} \right]^\dagger \, \mathcal{F}_{\text{fin}}^{(r)} \right\} \,.
\end{align}
Finally, we note that in the main text we have taken  $\ne=\nE=n_l/2$.

\begin{thebibliography}{99}

\bibitem{Sudakov:1954sw}
  V.~V.~Sudakov,
  Sov.\ Phys.\ JETP {\bf 3}, 65 (1956)
  [Zh.\ Eksp.\ Teor.\ Fiz.\  {\bf 30}, 87 (1956)].


\bibitem{Gorshkov:1966ht}
  V.~G.~Gorshkov, V.~N.~Gribov, L.~N.~Lipatov and G.~V.~Frolov,
  Sov.\ J.\ Nucl.\ Phys.\  {\bf 6}, 95 (1968)
  [Yad.\ Fiz.\  {\bf 6}, 129 (1967)].

\bibitem{Cheng:1970xm}
  H.~Cheng and T.~T.~Wu,
  Phys.\ Rev.\ D {\bf 1}, 2775 (1970).

\bibitem{Frolov:1970ij}
  G.~V.~Frolov, V.~N.~Gribov and L.~N.~Lipatov,
  Phys.\ Lett.\  {\bf 31B}, 34 (1970).


\bibitem{Frenkel:1976bj}
  J.~Frenkel and J.~C.~Taylor,
  Nucl.\ Phys.\ B {\bf 116}, 185 (1976).

\bibitem{Libby:1978qf}
S.~B.~Libby and G.~F.~Sterman,
Phys. Rev. D \textbf{18}, 3252 (1978).

\bibitem{Mueller:1979ih}
  A.~H.~Mueller,
  Phys.\ Rev.\ D {\bf 20}, 2037 (1979).

\bibitem{Collins:1980ih}
  J.~C.~Collins,
  Phys.\ Rev.\ D {\bf 22}, 1478 (1980).

\bibitem{Sen:1981sd}
  A.~Sen,
  Phys.\ Rev.\ D {\bf 24}, 3281 (1981).

\bibitem{Collins:1985ue}
J.~C.~Collins, D.~E.~Soper and G.~F.~Sterman,
Nucl. Phys. B \textbf{261}, 104 (1985).

\bibitem{Sterman:1986aj}
  G.~F.~Sterman,
  Nucl.\ Phys.\ B {\bf 281}, 310 (1987).


\bibitem{Korchemsky:1988hd}
  G.~P.~Korchemsky,
  Phys.\ Lett.\ B {\bf 220}, 629 (1989).


\bibitem{Laenen:2010uz}
E.~Laenen, L.~Magnea, G.~Stavenga and C.~D.~White,
JHEP \textbf{01}, 141 (2011).

\bibitem{Penin:2014msa}
  A.~A.~Penin,
  Phys.\ Lett.\ B {\bf 745}, 69 (2015), Erratum: [Phys.\ Lett.\ B {\bf 771}, 633
(2017)].

\bibitem{Melnikov:2016emg}
K.~Melnikov and A.~Penin,
JHEP \textbf{05}, 172 (2016).

\bibitem{Bonocore:2016awd}
D.~Bonocore, E.~Laenen, L.~Magnea, L.~Vernazza and C.~D.~White,
JHEP \textbf{12}, 121 (2016).

\bibitem{Liu:2017vkm}
  T.~Liu and A.~A.~Penin,
  Phys.\ Rev.\ Lett.\  {\bf 119},  262001 (2017).


\bibitem{Boughezal:2016zws}
R.~Boughezal, X.~Liu and F.~Petriello,
JHEP \textbf{03}, 160 (2017).


  \bibitem{Moult:2018jjd}
  I.~Moult, I.~W.~Stewart, G.~Vita and H.~X.~Zhu,
  JHEP {\bf 1808}, 013 (2018).

\bibitem{Liu:2018czl}
  T.~Liu and A.~Penin,
  JHEP {\bf 1811}, 158 (2018).

\bibitem{Beneke:2018gvs}
  M.~Beneke, A.~Broggio, M.~Garny, S.~Jaskiewicz, R.~Szafron, L.~Vernazza and J.~Wang,
  JHEP {\bf 1903}, 043 (2019).

\bibitem{Ebert:2018gsn}
M.~A.~Ebert, I.~Moult, I.~W.~Stewart, F.~J.~Tackmann, G.~Vita and H.~X.~Zhu,
JHEP \textbf{04}, 123 (2019).

\bibitem{Bahjat-Abbas:2019fqa}
N.~Bahjat-Abbas, D.~Bonocore, J.~Sinninghe Damst{\'e}, E.~Laenen, L.~Magnea, L.~Vernazza and C.~D.~White,
JHEP \textbf{11}, 002 (2019).

\bibitem{Beneke:2019mua}
M.~Beneke, M.~Garny, S.~Jaskiewicz, R.~Szafron, L.~Vernazza and J.~Wang,
JHEP \textbf{01}, 094 (2020).

\bibitem{Moult:2019uhz}
I.~Moult, I.~W.~Stewart, G.~Vita and H.~X.~Zhu,
JHEP \textbf{05}, 089 (2020).

\bibitem{Liu:2019oav}
Z.~L.~Liu and M.~Neubert,
JHEP \textbf{04}, 033 (2020).


\bibitem{Anastasiou:2020vkr}
  C.~Anastasiou and A.~Penin,
  JHEP {\bf 2007}, 195 (2020).

\bibitem{Liu:2020wbn}
Z.~L.~Liu, B.~Mecaj, M.~Neubert and X.~Wang,
JHEP \textbf{01}, 077 (2021).


\bibitem{Liu:2021chn}
T.~Liu, S.~Modi and A.~A.~Penin,
JHEP \textbf{02}, 170 (2022).


\bibitem{Beneke:2022obx}
M.~Beneke, M.~Garny, S.~Jaskiewicz, J.~Strohm, R.~Szafron, L.~Vernazza and J.~Wang,
JHEP \textbf{07}, 144 (2022).

\bibitem{AH:2020iki}
A.~A H, P.~Mukherjee and V.~Ravindran,
Phys. Rev. D \textbf{105}, 094035 (2022).


\bibitem{Bell:2022ott}
  G.~Bell, P.~B\"oer and T.~Feldmann,
  JHEP \textbf{09}, 183 (2022).

\bibitem{Bell:2024bxg}
G.~Bell, P.~B{\"o}er, T.~Feldmann, D.~Horstmann and V.~Shtabovenko,
JHEP \textbf{09}, 098 (2025).


\bibitem{Liu:2022ajh}
Z.~L.~Liu, M.~Neubert, M.~Schnubel and X.~Wang,
JHEP \textbf{06}, 183 (2023).

\bibitem{Liu:2024tkc}
T.~Liu, A.~A.~Penin and A.~Rehman,
JHEP \textbf{04}, 031 (2024).


\bibitem{Penin:2019xql}
  A.~A.~Penin,
  JHEP {\bf 04}, 156 (2020).

\bibitem{Bonciani:2021okt}
R.~Bonciani, A.~Broggio, S.~Di Vita, A.~Ferroglia, M.~K.~Mandal,
P.~Mastrolia, L.~Mattiazzi, A.~Primo, J.~Ronca and U.~Schubert, \textit{et al.}
Phys. Rev. Lett. \textbf{128}, 022002 (2022).

\bibitem{Penin:2005kf}
  A.~A.~Penin,
  Phys.\ Rev.\ Lett.\  {\bf 95}, 010408 (2005).

\bibitem{Penin:2005eh}
  A.~A.~Penin,
  Nucl.\ Phys.\ B {\bf 734},  185 (2006).

\bibitem{Bonciani:2007eh}
  R.~Bonciani, A.~Ferroglia, and A.~A.~Penin,
  Phys.\ Rev.\ Lett.\  {\bf 100},  131601 (2008).


\bibitem{Beneke:1997zp}
  M.~Beneke and V.~A.~Smirnov,
  Nucl.\ Phys.\ B {\bf 522}, 321 (1998).

\bibitem{Smirnov:2002pj}
  V.~A.~Smirnov,
  {\it Applied asymptotic expansions in momenta and masses},
  Springer Tracts Mod.\ Phys.\  {\bf 177 } (2002)  1.


\bibitem{Kuhn:1999nn}
  J.~H.~Kuhn, A.~A.~Penin and V.~A.~Smirnov,
  Eur.\ Phys.\ J.\ C {\bf 17}, 97 (2000).

\bibitem{Penin:2016wiw}
  A.~A.~Penin and N.~Zerf,
  Phys.\ Lett.\ B {\bf 760}, 816 (2016),  Erratum: [Phys.\ Lett.\ B {\bf 771},
637 (2017)].

\bibitem{Gardi:2022khw}
E.~Gardi, F.~Herzog, S.~Jones, Y.~Ma and J.~Schlenk,
JHEP \textbf{07}, 197 (2023).

\bibitem{terHoeve:2023ehm}
J.~ter Hoeve, E.~Laenen, C.~Marinissen, L.~Vernazza and G.~Wang,
JHEP \textbf{02}, 024 (2024).

\bibitem{Ma:2023hrt}
Y.~Ma,
JHEP \textbf{09}, 197 (2024).



\bibitem{Gardi:2024axt}
E.~Gardi, F.~Herzog, S.~Jones and Y.~Ma,
JHEP \textbf{08}, 127 (2024).

\bibitem{Delto:2023kqv}
M.~Delto, C.~Duhr, L.~Tancredi and Y.~J.~Zhu,
Phys. Rev. Lett. \textbf{132},  231904 (2024)


\bibitem{Supp}
Supplemental material.

\bibitem{Becher:2009kw}
T.~Becher and M.~Neubert,
Phys. Rev. D \textbf{79} (2009), 125004
[erratum: Phys. Rev. D \textbf{80} (2009), 109901]

\bibitem{Ferroglia:2009ep}
A.~Ferroglia, M.~Neubert, B.~D.~Pecjak and L.~L.~Yang,
Phys. Rev. Lett. \textbf{103} (2009), 201601

\bibitem{Czakon:2020vql}
M.~L.~Czakon and M.~Niggetiedt,
JHEP \textbf{05}, 149 (2020).

\bibitem{Peraro:2020sfm}
T.~Peraro and L.~Tancredi,
Phys. Rev. D \textbf{103}, 054042 (2021).

\bibitem{tHooft:1972tcz}
G.~'t Hooft and M.~J.~G.~Veltman,
Nucl. Phys. B \textbf{44}, 189 (1972).

\bibitem{Yennie:1961ad}
D.~R.~Yennie, S.~C.~Frautschi and H.~Suura,
Annals Phys. \textbf{13}, 379 (1961).


\end{thebibliography}
\end{document}